\documentstyle[11pt]{article}

\newcommand\afour{
\textheight 8.5in
\textwidth 6in
\oddsidemargin 0pt
\topmargin -30pt
}

\newcommand\eq[1]{Eq.~(\ref{#1})}
\newcommand\eqs[2]{Eqs.~(\ref{#1}) and (\ref{#2})}
\newcommand\eqss[3]{Eqs.~(\ref{#1}), (\ref{#2}) and (\ref{#3})}

\newcommand\rfrac[2]{\left(\frac{#1}{#2}\right)}

\newcommand\ee{\end{equation}}
\newcommand\be{\begin{equation}}
\newcommand\eea{\end{eqnarray}}
\newcommand\bea{\begin{eqnarray}}

\newcommand\sunit{\,\mbox{s}}
\newcommand\TeV{\,\mbox{TeV}}
\newcommand\GeV{\,\mbox{GeV}}
\newcommand\MeV{\,\mbox{MeV}}
\newcommand\keV{\,\mbox{keV}}
\newcommand\eV{\,\mbox{eV}}
\newcommand\km{\,\mbox{km}}

\newcommand\Mpc{\,\mbox{Mpc}}

\newcommand\mone{^{-1}}
\newcommand\mtwo{^{-2}}
\newcommand\mthree{^{-3}}
\newcommand\mfour{^{-4}}
\newcommand\mhalf{^{-1/2}}
\newcommand\mthreehalf{^{-3/2}}

\newcommand\mquarter{^{-1/4}}
\newcommand\half{^{1/2}}
\newcommand\threehalf{^{3/2}}

\newcommand\mpl{m_{Pl}}

\newcommand\lsim{\mathrel{\rlap{\lower4pt\hbox{\hskip1pt$\sim$}}
    \raise1pt\hbox{$<$}}}
\newcommand\gsim{\mathrel{\rlap{\lower4pt\hbox{\hskip1pt$\sim$}}
    \raise1pt\hbox{$>$}}}

\newcommand\sub[1]{{\mbox{\scriptsize{#1}}}}
\newcommand{\fa}{{f_\ax}}
\newcommand{\ftw}{{\rfrac{\fa}{10^{12}\GeV}}}
\newcommand{\mtev}{{\rfrac{m_\sax}{1\TeV}}}
\newcommand{\alon}{{\rfrac{\alpha_c}{.1}}}
\newcommand{\st}{{\sub{st}}}
\newcommand{\ax}{{\sub{a}}}
\newcommand{\sax}{{\sub{sax}}}
\newcommand{\saxeq}{{\sub{saxeq}}}
\newcommand{\decoup}{{\sub{decoup}}}
\newcommand{\decay}{{\sub{decay}}}
\newcommand{\qcd}{{\sub{QCD}}}

\newcommand{\reh}{{\sub{reh}}}
\newcommand{\kmin}{k_\sub{min}}
\newcommand{\calp}{{\cal P}}
\newcommand{\call}{{\cal L}}

\afour

\begin{document}
\title{DILUTION OF COSMOLOGICAL DENSITIES \\BY SAXINO DECAY}
\author{David H.~Lyth\thanks{e-mail LYTH@UK.AC.LANCS.PH.V1}\\[.5em]
School of Physics and Materials,\\
Lancaster University,\\
Lancaster LA1 4YB, \\
U.K.}
\date{May 1993}

\maketitle

\begin{abstract}
Saxino decay can generate
significant cosmological entropy, and hence dilute theoretical estimates of
the present mass density of a given particle
species.  The dilution factor depends on the saxino and axion masses,
and is constrained by the requirement that saxino decay should not
affect nucleosynthesis, as well as by the usual requirement that the axion
density be less than the critical density. The latter constraint is evaluated
carefully, under both the Harari-Sikivie and Davis proposals about the
emission spectrum from axionic strings. Uncertainties are carefully evaluated,
points of principle are addressed, and with an eye to
future numerical simulation the spacing and typical oscillation
wavelength of the strings are represented by parameters varying in the
range 1 to 3. Within the constraints, the entropy dilution varies from
1 to $10^{-4}$. Only saxinos originating from thermal equilibrium
are considered, so that more dilution might arise from non-thermal saxinos.
\end{abstract}

\newpage

\section{Introduction}

According to current ideas, the present mass density $\Omega$ of the universe
is practically equal to 1 (in units of the critical density), baryons
contributing $\Omega_B\sim .04$ to .07,  and one or more species of dark matter
particle making  up the rest. Given a suitable model of the fundamental
interactions one can calculate the present density $\Omega_X$ of
a given stable particle species. The calculation usually proceeds in two
stages \cite{kotu90}. First one calculates the particle number density
at some initial epoch, defined by its temperature $T_i$, after which the
particle number is conserved.
Then the number density is evolved forward to the present
epoch when the (photon) temperature is $T_0=2.74$\,K.

As long as thermal equilibrium is maintained, the forward evolution
can be performed using entropy conservation. Entropy can however be
generated by non-equilibrium processes, and this lowers the prediction
for $\Omega_X$. A particular example is the
decay of a particle species which dominates the energy density of the
universe during some era prior to its decay.
The Standard Model does not contain such a particle species,
but its as yet unknown extension might.

Any acceptable extension must include a natural explanation of
CP conservation by the strong interaction. Although it has recently been
seen to be rather vulnerable to Planck scale corrections \cite{hoet92,kama92},
the most promising explanation is still a spontaneously broken global
symmetry of the type proposed
by Peccei and Quinn \cite{pequ77,kim87}. Along with
this symmetry comes a Goldstone boson, the axion
\cite{wein78,kim87,kotu90,lind90}. The axion is practically stable and
has extremely weak interactions, constituting cold dark matter.
{}From accelerator physics and astrophysics
its mass has an upper bound $m_\ax<10\mthree\eV$, and from the cosmological
requirement $\Omega_\ax<1$ its mass has a lower bound
$m_\ax>m_\sub{min}$. The lower bound $m_\sub{min}$
is difficult to calculate, but with
no entropy production it is probably not far below $10\mthree\eV$.
In that case there is only a narrow allowed window for $m_\ax$, and
the axion must make up a significant fraction of the dark matter.

Despite the  fact that no supersymmetric particle has yet been observed,
it is fair to say that an increasing number of particle theorists
feel that an acceptable extension of the Standard
Model should also respect (low energy) supersymmetry \cite{wilc93}.
 One reason for this feeling is
the successful prediction of $\sin^2\theta_W$ by supersymmetric
grand unified models, and another is the fact that the rival
technicolour models are coming under pressure
from ever more accurate measurements, notably at CERN.
Supersymmetry requires
\cite{nill84}  that the known particle species have
as yet undiscovered superpartners,
with masses of order $.1$ to $1\TeV$.
It also requires that the axion be
accompanied by a spin
$1/2$ partner called the axino, and a spin $0$ partner called the saxino
\cite{kim84,first,kim91}.
The axino might have the typical mass mentioned above, in which case it
would decay (perhaps with significant entropy production),
or it might be much lighter in which case it could constitute `warm'
dark matter. (More complicated possibilities arise if there is a very
light gravitino \cite{chki93}, which are discounted in the present paper.)
The saxino, on the other hand, is definitely
expected to have a mass in the usual range,
$100\GeV\lsim m_\sax \lsim 1\TeV $.

Kim \cite{kim91} has pointed out that
saxino decay could generate significant entropy,
and in the present paper the amount of entropy production
is carefully calculated as a function of the axion mass $m_\ax$, the
saxino mass $m_\sax$ and the maximum temperature after inflation
(`reheat' temperature) $T_\reh$. Then,
the cosmologically allowed region of this three parameter space is
delineated, by the requirement that saxino decay must not interfere
with nucleosynthesis, as well as by the requirement that the axion density
satisfies $\Omega_\ax<1$. As described in the conclusion, the results
have a number of cosmological implications.

Throughout the
units are $\hbar=c=1$, $\mpl=1.2\times10^{19}\GeV$ is the Planck mass,
 $a(t)$ is the scale factor of
the universe at time $t$, $H=\dot a/a$ is the Hubble parameter,
and $H_0=100\km\sunit\mone\Mpc\mone h$ is its present value.

\section{Entropy and particle densities}

For reference we recall some basic facts about radiation (relativistic
particles) in the early universe
\cite{kotu90}. According to the Standard Model the radiation is in thermal
equilibrium when the temperature $T$ exceeds $1\MeV$, with practically
zero chemical potential. Accordingly its density is
\be
\rho=\frac{\pi^2}{30} g_* T^4 \label{2}
\ee
where $g_*$ is the effective number of relativistic particle species
Its entropy density $s$ is
\be
s=\frac43 \frac{\rho_r}{T}=\frac{2\pi^2}{45} g_* T^3 \label{3}
\ee
With three massless neutrino species,
$g_*=10.75$ for $1\MeV\lsim T\lsim 100\MeV$, $g_*=17.25$ for
$100\MeV\lsim T\lsim\Lambda_\qcd$ and $g_*=61.75$ for
$\Lambda_\qcd\lsim T\lsim 2\GeV$  where $\Lambda_\qcd\simeq200\MeV$.
At higher temperatures
the Standard Model predicts a leveling out to $g_*\sim 100$ above
$T\sim 100\GeV$. However, extensions of the Standard
Model have additional particles, leading to a bigger value of $g_*$,
and in the minimal supersymmetric extension $g_*$ levels out at $g_*=229$.
Extending
the Standard Model does not usually have any other effect on
the formulas that we are reviewing in this section, except for the
possibility of entropy production that is our central concern.

In thermal equilibrium the entropy
$S\propto a^3 s$ in a comoving volume is constant. This implies that
\be a^3 g_*(T) T^3=\mbox{constant} \ee
After the neutrinos decouple at $T\sim1\MeV$ they are not in thermal
equilibrium but their entropy continues to be conserved, and the same is
true of the photons when they too decouple at a much later epoch.
Thus, according to the Standard Model the total entropy in a comoving volume is
conserved, and a standard calculation \cite{kotu90}
gives the present entropy density
$s_0=2938\,\mbox{cm}\mthree $.
This allows one to calculate the present mass density of any
particle species $X$, whose number density is conserved after some epoch
$T_i>1\MeV$, given the number density at that epoch. Indeed, in units of
the present critical density $3H_0^2\mpl^2/8\pi$ it is obviously
given by
\bea
\Omega_X&=& \frac{8\pi}{3}\frac{m s_0}{\mpl^2H_0^2} \frac{n_i}{s_i}\\
&=& 2.44 h\mtwo \frac{m}{10\eV} \frac{n_i}{s_i}
\label{6}
\eea
This relation has been used to calculate the baryon density
$\Omega_B$ and the density $\Omega_X$ of various dark matter candidates,
in terms of theoretically calculated initial number densities.

The above analysis may fail if there is a non-relativistic
particle species with a lifetime
in the range $t_i<\tau<t_0$. However, the required modification
is very simple provided that the decay products thermalise quickly
on the Hubble timescale. Then the radiation density and entropy density
are still given by \eqs{2}{3}, and the only effect of the decay is to
reduce $\Omega_X$ by a factor
$\Delta=S_0/S_i$, the increase in the entropy of a comoving volume
\be
\Omega_X=2.44 h\mtwo \frac{m}{10\eV} \frac{n_i}{s_i} \Delta\mone
\label{7}
\ee
Neither the Standard Model or its minimal supersymmetric extension
contains such a particle species, but for extensions which include
Peccei-Quinn symmetry the saxino can be such a species, as we now discuss.

\section{Entropy production from saxino decay}

The supersymmetric generalisation of the usual axion-gluon interaction
is  \cite{kim91}
\be
\call= \frac{\alpha_c}{16\pi\fa} \Phi W_i W^i
\ee
where the superfields are
$\Phi$ corresponding to the
axion and saxino (spin 0) and the axino (spin $1/2$), and
$W_i$ corresponding to the gluon (spin 1) and the
gluino (spin $1/2$). This reproduces the usual axion-gluon interaction,
and also gives a saxino-gluon interaction and other interactions,
\be {\cal L}=\frac{\alpha_c}{8\pi\fa} ( \psi_\ax G_{\mu\nu} \tilde
G^{\mu\nu}+ \psi_\sax G_{\mu\nu} G^{\mu\nu} + \dots) \ee
The saxino is kept in thermal equilibrium by reactions like
$q\bar q\leftrightarrow sg$ and $gg\leftrightarrow sg$ above a
temperature \cite{kim91}
\be T_\decoup=10^{11}\GeV \ftw^2 \alon\mthree \ee
At an energy scale of order $10^{11}\GeV$, $\alpha_c\simeq 1/20$
in the minimal\footnote
{Since Peccei-Quinn symmetry is assumed the extension cannot in fact be
minimal, but this estimate of $\alpha_c$ and the estimates of $g_*$
used later will hopefully still be adequate.}
supersymmetric standard model
\cite{amet91},
so that
\be T_\decoup\simeq 8\times 10^{11}\GeV \ftw^2  \ee
After
decoupling the saxino number density $n_\sax$ is given as a fraction
of the entropy density by
\be r\equiv \frac{n_\sax}{s}=\frac1{3.6 g_*} =1.2\times 10\mthree \ee
(setting $g_*=229$ corresponding to the minimal
supersymmetric standard model).

In order to achieve thermal equilibrium the reheat  temperature
must satisfy $T_\reh>T_\decoup$. Unfortunately, even the order of
magnitude of $T_\reh$ is unknown.
Bounds on the cmb anisotropy, as
well as the COBE detection, imply that it is less than
$10^{16}\GeV$
\cite{lyth84,lily93a}.
 If there is a gravitino with mass $m_g<1\TeV$
nucleosynthesis implies that it is less than
$10^{13}\GeV$, but that bound rapidly
goes away as $m_g$ is increased and provides no additional constraint
if $m_g\sim 10\TeV$ \cite{kasa87}.

 In what follows we leave $T_\reh$ as a free parameter,
and assume that if $T_\reh<T_\decoup$ the saxino density is negligible,
considering only saxinos of thermal origin. A significant non-thermal saxino
density might originate as an inflationary fluctuation or be radiated
from axionic strings, possibilities which should certainly be
investigated.

Continuing the story of the thermal saxinos,
they become non-relativistic at a temperature $T\sim m_\sax$,
and if they live long enough
they dominate the energy density of the
universe below the temperature
\be T_\saxeq=\frac43 rm_\sax=1.6\GeV \mtev \ee
(The subscript `saxeq' indicates that the saxino density is equal to
that of the radiation at this temperature.)
The dominant decay mode of the saxino is
into two gluons, with lifetime $\Gamma\mone$ given by \cite{kim91}
\be \Gamma=\frac{\alpha_c^2 m_\sax^3}{128\pi^3\fa^2} \ee
and saxino domination will occur for a significant period if
$t_\saxeq\ll \Gamma\mone$.

With this assumption, the evolution of the matter and radiation densities can
easily be worked out. Approximate
analytic results are available
on the assumption that $g_*$ is constant
\cite{lapa86,lasc90,sctu85,sctu88,kotu90}, which are now briefly recalled.
The temperature corresponding to $t_\decay$ is
\be
T_\decay=.55 g_*\mquarter \sqrt{\Gamma m_P}
\ee
and the saxino dominates the energy density until this epoch.
The density of the `new' radiation from saxino decay becomes equal to
that of the `old' radiation at a temperature $T_=$, given by
\be T_=^5= T_\decay^4 T_\saxeq
\label{17}
\ee
Before this epoch there is no significant entropy generation, but after
it the entropy in a comoving volume is proportional to $T^{-5}$.
This continues until the epoch $T_\decay$, after which the saxino
density becomes negligible and entropy production stops.
Thus, if $T_i<T_=$ the entropy generation factor in \eq{7} is
\be
\Delta=\rfrac{T_i}{T_\decay}^5 \hspace{4em} (T_i<T_=)
\label{17a}
\ee
It depends strongly on $T_i$ because only part of the entropy is
generated after $T_i$.
On the other hand, if $T_i>T_=$,
all of the entropy is generated after $T_i$, and $\Delta$ is given by
\be
\Delta_\sub{total}=
\rfrac{T_=}{T_\decay}^5=\frac{T_\saxeq}{T_\decay}
\hspace{4em} (T_i>T_=)
\label{18a}
\ee
The last equality follows from \eq{17}.
It implies incidentally that the new radiation has caused the
ratio $\rho_\sax/\rho_r$ to fall to a value $\simeq1$,
just before it falls off exponentially at the epoch $T_\decay$.

These results are valid if $g_*$ is constant during the interval
$T_\decay<T<T_\saxeq$. Setting $g_*=10.75$, appropriate to the range
$1\MeV<T<100\MeV$, gives
\be
T_\decay=52\MeV \ftw \mone \mtev\threehalf
\label{19}
\ee

The three temperatures
$T_\decay<T_=<T_\saxeq$ are plotted in Figure 1.
If the saxino is heavy, $T_\decay$ can exceed
the temperature $T=\Lambda_\qcd$ at which $g_*$ rises sharply to
62, but using this value would not change the results much.
We have not investigated the intermediate case
$T_\decay<\Lambda_\qcd<T_\saxeq$, but assume that the results above are
still valid to sufficient accuracy.

Substituting \eq{19} into \eq{18a} gives the maximum
entropy generation factor which can be supplied by saxino decay,
\be
\Delta_\sub{total} = 31 \ftw \mtev\mhalf
\label{20}
\ee

The saxino
must decay before it has a significant effect on nucleosynthesis.
According to Scherrer and Turner \cite{sctu88} a safe bound is
$t<10\sunit$, which is
equivalent to
$T_\decay\gsim.3\MeV$, or
\be
\ftw\lsim 170\mtev\threehalf
\ee
The region forbidden by this constraint is indicated in
Figure 1. The corresponding constraint on the entropy generation factor
is
\be
\Delta_\sub{total}< 5.2\times 10^3 \mtev
\ee
A further constraint can arise from the requirement $\Omega_\ax<1$,
which is our main concern for the rest of the paper.
as is now investigated.

\section{The axion density}

The axion density has been the subject of numerous investigations.
The situation as it was
understood in 1989 is described in \cite{kotu90,lind90}, and
further work has been done since then
\cite{lyth90,lily90,tuwi91,lind91,lyth92a,lyth92b,lyst92,kama92}.
There are essentially two possibilities regarding axion cosmology.
One is that the Peccei-Quinn field has been in its vacuum
(ie., that PQ symmetry has been spontaneously broken) ever since the
observable universe left the horizon during inflation. In that case
axion cosmology is quite complicated as will be briefly recalled in
Section 4.3.

More likely is the opposite case, where
Peccei-Quinn symmetry is restored in the early universe, through either
finite temperature effects (after inflation) or the quantum fluctuation
(during inflation). In that case
strings form at the epoch when the symmetry is spontaneously broken,
and the strings oscillate until domain walls form when the axion mass
switches on at $T\simeq 1\GeV$. If the string network `scales' as one
expects, $\Omega_\ax$ can in principle be calculated uniquely in terms of
$f_\ax$ (up to the entropy generation factor $\Delta\mone$), at least if
the strings radiate nothing but axions.
To do a proper calculation requires numerical simulations which
are perhaps tractable while the strings are oscillating but which would
be very difficult after domain walls form. In the absence of such
simulations, $\Omega_\ax$ has been estimated in the literature on the
basis of simplifying assumptions. The first of these, used by
essentially all investigators, is that axion number is conserved after
the domain walls form, so that \eq{7} gives $\Omega_\ax$ in terms of the
axion number density just before wall formation.
The second assumption is a statement about the shape of the
spectrum of axions emitted by the strings.
Two alternative proposals exist,
one by
 Davis \cite{davi86} and one by
Harari and Sikivie \cite{hasi87}.
In what follows these two proposals are critically
assessed, and in each case $\Omega_\ax$ is carefully estimated.
The treatment represents an advance on previous ones
(except for \cite{lyth92a} which is
discussed below) in two respects. First,
the spectrum of {\em all} axions present at a given
epoch is calculated (not just that of those being emitted), which
enables important points of principle to be addressed.
Second, the result is given as a function of
the string spacing, as well as (for Davis' proposal)
the string oscillation frequency. Correspondingly there appear in the result
two parameters
$\gamma$ and $\beta$, which are expected to lie in the range
$1$ to a few, and which should be calculable in the forseeable future
from numerical simulations.
Pending such a calculation, we estimate  $\Omega_\ax$
by allowing them to vary between 1 and 3.

\subsection{Axion cosmology}
First we need some basic
properties of the axion
\cite{kim87,kotu90}. It is a
practically stable particle, which is massless for
$T\gg \Lambda_\qcd\simeq 200\MeV$. At lower temperatures its
interaction with the gluon
leads to an effective potential
\be
V=(79\MeV)^4 (1-\cos\theta)
\label{22}
\ee
where $\theta=\psi_\ax/f_\ax $ and $\psi_\ax$ is the canonically
normalised axion field. (We ignore a possible generalisation
of the above equations, $\theta\to N\theta$ and $\fa\to\fa/N$
with $N$ an integer, because it probably
leads to cosmologically unacceptable domain walls.)
Small oscillations around the minimum of this potential correspond to free
axions with mass $m_\ax$ given by $f_\ax m_\ax=79\MeV$, or
\be
\frac{m_\ax}{10^{-6}\eV }=6.2 \ftw\mone
\ee
Accelerator physics and astrophysics provide a lower limit
$m_\ax\lsim 10\mthree \eV$, corresponding to $f_\ax\gsim 10^{10}\GeV$
\cite{rafe90}.

\subsubsection*{The case of a homogeneous axion field}

As already mentioned, axion cosmology depends crucially on whether or
not there are axionic strings.
Before turning to the former case which is our main concern, we focus on
the simplest possible example of the latter case. Namely, we assume that
the axion field is spatial homogeneous.
In this case an accurate estimate of $\Omega_\ax$ is provided by
the calculation of Turner \cite{turn86},
which we briefly recall. It will form the basis of our later
estimates for the more complicated string case.

The axion mass switches on gradually as the epoch $T=\Lambda_\qcd$,
becoming
significant at the epoch $\tilde t$ defined by
\be
m_\ax(\tilde t)=3H(\tilde t)
\label{24}
\ee
Before this epoch, the homogeneous axion field is time independent,
and afterwards it oscillates around $\theta=0$. If $|\tilde\theta|
\ll \pi$
(a tilde on any quantity will always denote its value at the epoch
$\tilde t$),
the oscillations are harmonic with angular frequency
$\tilde m$ and amplitude proportional to $a\mthreehalf
m_\ax\mhalf$, corresponding to
the presence of zero momentum non-interacting axions with number
density
\be
\tilde n=\frac12 \tilde m \fa^2 \tilde \theta ^2
\label{25}
\ee
(Note that $n\propto a\mthree$ corresponding to conserved axion
number \cite{absi83}.) The present axion density
is therefore given by \eq{6} once $\tilde T$ is known. While the axion mass
is switching on it is given by
\be
m_\ax(T)/m_\ax=.077 (\Lambda_\qcd/T)^{3.7}
\label{temp}
\ee
Taking $h=.5\pm.1$ and $\Lambda_\qcd=(200\pm50)\MeV$ this leads to
\be \tilde T=.87 \ftw ^{-.18} \GeV
\label{26}
\ee
and
\be
\Omega_\ax=0.9\times 10^{\pm.5}
\ftw^{1.18} \Delta\mone \tilde\theta^2
\label{27}
\ee
The quoted uncertainty is essentially the one evaluate by Turner as
arising from the uncertainties in
$g_*$ and $m_\ax(T)$, slightly increased to allow for
additional uncertainty arising from the values of $h$ and
$\Lambda_\qcd$.
(Bearing in mind the age of the universe
we here discount the possibility $h\simeq1$, which according to
\eq{7} would multiply $\Omega_\ax$ by a factor $.25$.)
To this level of accuracy, the expression is valid  for
$|\tilde\theta|\lsim.9\pi$, only
the case $|\tilde \theta|\simeq \pi$ requiring special treatment
\cite{turn86,lyth92b}.

The entropy production factor $\Delta$ is equal to 1 if $T_\reh<T_\decoup$
or if $T_\saxeq<T_\decay$ where the four temperatures are defined in the
last section. Otherwise it is given by \eqs{17a}{18a}, with
$T_i=\tilde T$. From the left hand column of Figure 1, one sees that
unless the saxino is rather heavy $\tilde T$ is bigger than $T_=$, so that
the total entropy generation factor is experienced by the axions,
given by \eq{20}.

\eqs{26}{27} are derived under the
standard assumption that
$\tilde T$ occurs during radiation domination. From Figure 1 one sees that
if the saxino is rather heavy $\tilde T$ can occur during
(saxino) matter domination, in which case there are well defined
correction factors \cite{lasc90}. In the present context these are not
very significant, and will be ignored.

The expression for $\Omega_\ax$ given by \eqss{17a}{18a}{27}
does not agree with that
of Kim \cite{kim91}.
 The reason is that he used formulae developed by Lazarides
{\em et al} \cite{lasc90} in a different context, where
the decaying object is not a saxino and where
$\tilde T$ falls in the interval $T_\decay<T<T_=$ throughout the relevant
portion of parameter space.
Figure 1 illustrates that this is far from true in the present
case.

\subsection{The string scenario}

The above calculation provides a basis for calculating
$\Omega_\ax$ on the assumption that there are axionic strings,
if axion number is conserved after domain walls form
(the validity of this assumption is briefly addressed at the end of the
 present section). By looking at the typical spatial and temporal
variation of $\theta$ it was argued in \cite{lyth92a} that the walls
form at about the temperature $\simeq\tilde T$, and this estimate will be
used
in what follows. (More precisely it was argued that the formation
temperature is lower by an insignificant factor $\gamma^{-1/6}$
where $\gamma$ is the string spacing introduced below.)
Substituting \eq{25} into \eq{27} gives an expression for $\Omega_\ax$
in terms of the number density $\tilde n$ just before wall formation.
\be
\Omega_\ax=0.9\times 10^{\pm.5} \ftw^{1.18}
\left( 2\tilde n \tilde m_\ax\mtwo f_\ax\mtwo \right)
\Delta\mone
\label{29}
\ee

To estimate $\tilde n$, the
simplest assumption is that
$\theta$ is as homogeneous as it can be,
subject to the requirement that it changes by $2\pi$
as one goes around a string.
This implies that the typical value of $\theta$ is of order 1 radian,
and to a rough approximation it also implies that
the space and time dependence of $\theta$
can be ignored in the equation of motion.
After wall formation there is an a roughly homogeneous axion field
oscillating with initial amplitude $\tilde\theta\sim1$,
giving $\tilde n\simeq \frac12 \tilde m_\ax^2 f_\ax^2$ and
\be
\Omega_\ax\sim 1\times \ftw^{1.18} \Delta\mone
\label{29a}
\ee
As we now discuss, this simplest estimate was queried by
Davis \cite{davi86}, who argued that axion radiation from strings plays
a crucial role.

\subsubsection*{Energy loss from the string network}

Assume that there is a network of axionic strings,
with at least
one piece of string passing through a typical Hubble volume.
On scales bigger than the horizon the string network expands with the
universe, and on smaller scales string annihilation occurs.
We assume that the string network has the `scaling'
property, whereby it looks the same at each epoch when viewed on the
Hubble scale. To be precise, we assume that the energy density of the
string network can be written
\be
\rho_\st=H^2\mu\gamma^2
\label{28}
\ee
where $\mu$ is the string energy per unit length and $\gamma$ is time
independent. Roughly speaking, $\gamma$ is the string spacing in units of the
Hubble distance $H\mone$.

The energy per unit length $\mu$ is dominated by the axion field
around the string, rather than by the string core.
The latter
has thickness $\fa\mone$ and energy per unit length $\fa^2$,
but a static axion field out to a distance $R$ gives energy per unit
length $2\pi\fa^2 \ln(R\fa)$. As we shall discuss in a moment the axion
field far away from the string is oscillating, but as a rough estimate
we can set $R$ equal to the string spacing $(H\gamma)\mone$.
Then $\mu$ has only logarithmic time dependence, and at
the relevant epoch $T\sim 1\GeV$ it is given by $\mu=2\pi\fa^2\eta$
where $\eta=\ln(\fa/\gamma H)$.
With $\fa\sim 10^{12}\GeV$ and $\gamma\sim1$ this gives the estimate
 $\eta\simeq70$, which  we
use in what follows.

{}From \eq{28} one can calculate the rate at which the string network
loses energy \cite{davi86}.
As in the homogeneous case we assume radiation domination throughout
the string oscillation era. One then has
$H=1/2t$ and \eq{28} becomes
\be
\rho_\st=\frac\pi 2 \gamma^2 \fa^2 \eta t\mtwo
\ee
The condition that comoving strings lose no energy is
$\rho_\st\propto a\mtwo$. To maintain the scaling solution, energy
conservation therefore requires that a comoving volume
$a^3$ of the string network emits energy
at a rate $a^3 R$ where $R$ is given by
\be
R=-a\mtwo\frac{d (a^2\rho_\st)}{dt}=\frac\pi 2 \gamma^2 \fa \eta t\mthree
\ee
All of this energy is assumed to be gained by the axion field.
In the absence of the string network this field would correspond to a
collection of massless non-interacting axions with energy density
$\rho_\ax\propto a\mfour$, so energy conservation requires
\be
a\mfour \frac{d (a^4\rho_\ax)}{dt} =R
=\frac\pi 2 \gamma^2 \fa \eta t\mthree
\label{31}
\ee

\subsubsection*{Estimating the axion number density}

So far everything is rigorous, given the scaling assumption. To go
further one has to introduce the concept of an `axion' with well defined
momentum and energy, as opposed to
just an `axion field'. Such an axion corresponds to field configuration
which is a plane wave, and this clearly requires that the wavelength is
much less than the string spacing, or equivalently that the wavenumber
$\omega=k/a$ is much bigger than
\be
\omega_\sub{min}\equiv\frac{\kmin}{a}=2\pi\gamma H
\label{38}
\ee

For a plane wave with amplitude $\theta$, the energy density
$\rho_\ax$ is related to the number density $n$ by $\rho_\ax=\omega n
_\ax$
where $\omega=k/a$ is the angular frequency.
Let us define the spectrum
$\calp_\ax$ of the axion energy density as the contribution
to it from unit interval of $\ln k$,
\be \rho_\ax=\int^\infty_{\kmin} \calp_\ax(t,k) \frac{dk}{k} \ee
then
\be n=\int^\infty_{\kmin} \calp_\ax(t,k) \rfrac ak \frac{dk}{k}
\label{37}
\ee

One can also define the
spectrum of the axions which are
being emitted during some small time interval $dt$,
which we write as $dt \calp_\sub{emis}$.
It is related  by energy-momentum conservation to the axion
spectrum (cf. \eq{31}),
\be
\calp_\sub{emis}(t,k)=a\mfour \frac{\partial}{\partial t}
\left( a^4 \calp_\ax(t,k) \right)
\label{39}
\ee
Integrating over $k$ gives the energy conservation constraint \eq{31},
\be
\int^\infty_{k_\sub{min}}
\calp_\sub{emis}(t,k)\frac{dk}{k} =\frac\pi2 \gamma^2 \fa^2 \eta t\mthree
\label{40}
\ee
Given a hypothesis about its shape, this equation determines
$\calp_\sub{emis}$, and integrating \eq{39} then gives
$\calp_\ax$ which gives the
number density through \eq{37}.

Two proposals have been made about the shape of the spectrum.
According to Davis \cite{davi86,dash89}, axion
emission is caused by smooth
oscillations of the string, with wavelength roughly $\beta\mone$ times
the string spacing, where $\beta$ is between 1 and a few.
This corresponds to
\be \calp_\sub{emis}(t,k)=f(t) \delta(k-k_*(t)) \label{41} \ee
where
\be \omega_*\equiv\frac{k_*}a= \beta \omega_\sub{min}
=\frac{\pi \gamma\beta}{t}
\label{42}
\ee
Inserting this expression into \eq{39} gives
\be \calp_\ax(t,k)=\pi\eta \gamma^2 \fa^2 t\mtwo \theta(k-k_*(t)) \ee
and then \eq{37} gives
\bea n&=&\pi \eta\gamma^2 f_\ax^2 t\mtwo \int^\infty_{\beta\omega_\sub{min}}
\frac{d\omega}{\omega^2}\label{44}\\
&=& \frac{\gamma\eta}{\beta}f_\ax^2 t\mone
\eea
leading to
\be \Omega_\ax=1.2\times10^{\pm.5}\ftw^{1.18} \gamma \eta\beta\mone
 \Delta\mone
\label{45}
\ee
This is a factor of order $\gamma\eta\beta\mone$ times
the simple estimate \eq{29a}. Taking for definiteness
$1<\gamma<3$ and $1<\beta<3$ leads to
\be
\Omega_\ax=(9\mbox{ to }800)\ftw^{1.18}  \Delta\mone
\ee
In other words, taking all of the uncertainties into account Davis'
proposal increases the simple estimate by a factor
of order 10 to 1000.

The proposal of Harari and Sikivie \cite{hasi87,hasi90} is that
the emission spectrum is flat,
\be \calp_\sub{emis}(t,k)=f(t) \hspace{5mm} (\frac{\kmin}a<\frac ka < \fa )\ee
This gives
\bea \calp_\ax(t,k)&=&\pi\gamma^2\fa^2t\mtwo\ln\rfrac{k}{\kmin(t)} \\
n&=&\pi \gamma^2 f_\ax^2 t\mtwo \int^\infty_{\omega_\sub{min}}
\ln\rfrac{\omega}{\omega_\sub{min}} \frac{d\omega}{\omega^2}
\label{48}\\
&=& \fa^2 \gamma t\mone \\
\Omega_\ax&=& 1.2\times10^{\pm.5}\ftw^{1.18} \gamma \Delta\mone
\label{49}
\eea
This is only a factor or order $\gamma$ times the simple estimate.
Taking $1<\gamma<3$ it corresponds to
\be
\Omega_\ax=(0.4\mbox{ to }11)\ftw^{1.18}  \Delta\mone
\ee

In the two right hand columns of Figure 1 are shown the bands
which correspond to these estimates. The full lines correspond to
$T_\sub{reh}=10^{16}\GeV$, the biggest possible value. The dotted lines
(and their straight continuation) correspond to the absence of
supersymmetry, or equivalently to $T_\sub{reh}<10^9\GeV$.
The allowed window for $m_\ax$ is shown in the two left hand columns of
Figure 2 as a function of $T_\reh$.

\subsubsection*{Critical assessment of the proposals}

According to both estimates,
most of the axions have angular frequency not far above $\omega_\sub{min
}$ (\eqs{44}{48}). In other words the condition that their wavelength is
much less than the string spacing, necessary for the axion concept to
make sense, is not terribly well satisfied.
Indeed, if $\gamma=1$ and (in the Davis case) $\beta=1$,
it is at best marginally satisfied; most
axions then have wavelength of order the string spacing. In that case,
one might ask why the rough estimate \eq{29a} is not reproduced, which
after all started with precisely the assumption that the wavelength was
of order the string spacing. A clue to the answer appears when one
calculates the mean square axion field
\cite{lyth92a},
\be \fa^2 \overline{\theta^2 }=\int^\infty_{\kmin} \calp_\ax(t,k)
\rfrac ak^2 \frac{dk}{k} \ee
(which follows from the fact that a relativistic
plane wave with amplitude $\theta$
and frequency $\omega$ has energy density $\fa^2\theta^2\omega^2$).
The corresponding mean square amplitude of the oscillation
in $\theta$ is
\bea \overline{\theta^2}&=& \frac\eta{2\pi\beta^2}\simeq
\rfrac{3.3}{\beta}^2 \mbox{\hspace{5mm} (Davis)}\\
\overline{\theta^2}&=& \frac1{2\pi}\simeq
(.40)^2 \mbox{\hspace{5mm} (Harari-Sikivie)} \eea

If we set $\tilde\theta^2=\overline{\theta^2 }$ in \eq{27}
then \eqs{45}{49} are roughly
reproduced for the the case $\beta=\gamma=1$
(to be precise the result is a factor $3/8\pi$ smaller).
In other words, the essential reason why the
Harari-Sikivie proposal reproduces the simple estimate
is that it leads to an amplitude $\theta\sim1$, and the essential reason
why the Davis proposal is bigger by a factor of order $\eta$ is that it leads
to an amplitude  $\theta\sim\eta\half$.

This last fact is rather worrying.
If $\eta$ had been a few
orders of magnitude bigger than $70$ the
amplitude would have been much bigger than $2\pi$.
Such an oscillation amplitude
would not in itself be unphysical, but the problem would be that it
cannot be generated by the smooth,
only marginally relativistic string oscillations envisaged in Davis'
proposal; rather, such oscillations generate a typical amplitude
$\theta\sim 1$ \cite{dash89}.
Thus Davis' proposal would by inconsistent with
the scaling assumption if $\eta$ were much bigger than its actual value.
One wonders what mechanism resolves the conflict
when it arises, and whether this mechanism is already beginning to
operate in the marginal case that we are dealing with.\footnote
{The above viewpoint is somewhat different from the one taken by the
present author in \cite{lyth92a}. There, a less careful calculation
gave a somewhat larger amplitude. More importantly, the fact
that Davis' string oscillations cannot generate an amplitude
$\theta\gg2\pi$ was
not appreciated, and an estimate was made of the axion density resulting
from such an amplitude. While this estimate would indeed be valid
if such an amplitude were somehow generated, it does not apply to
the the case at hand. I am indebted to Rick Davis for a clarifying
discussion on this question.}

Nevertheless, the statement that the oscillating strings
emit radiation with a wavelength of order their spacing
seems rather natural and to this extent the proposal of Davis is
perhaps a reasonable working hypothesis. By contrast
the proposal of Harari and Sikivie
is extremely radical, in that
it postulates the existence of a mechanism whereby the strings emit radiation
impartially with all wavelengths, right down to the string thickness
$\fa\mone\lsim 10^{-10}\GeV\mone$, no matter how late the epoch.
As the string thickness is only $10^{-28}$ of the string spacing by the
time radiation finishes, one would like to know something about the mechanism
before taking the proposal too seriously

As only strings as opposed to domain walls are involved it would seem
feasible to perform numerical simulations. They would hopefully confirm
the scaling behaviour, and determine the string spacing $\gamma$
defined by \eq{28} and emission spectrum, thus allowing one to calculate
$\Omega_\ax$ on the assumption that axion number is conserved after wall
formation. We end this section by briefly asking about the validity of
that assumption.

\subsubsection*{After domain wall formation}

The string-wall network must annihilate well before the present to avoid
cosmological disaster, and in order to make this possible we have
assumed that the vacua on the two sides of a wall are
identical (\eq{22}). How does annihilation proceed?
Based on
\cite{kila82}, the standard assumption in the literature seems to be
that gravitational radiation is the only significant process.
The underlying idea, inspired by what happens
for a loop of gauge string, is that the network has significant
structure only on macroscopic scales, of order the Hubble distance and bigger.
However, the walls are made out of the axion field and have a
thickness of order $1/m_\ax$, and it therefore
seems clear that collisions between
pieces of wall will result in the emission of marginally relativistic
axions at some level. Since gravity is so weak, this process is
presumably the dominant one.
As discussed in \cite{lyth92a}
wall annihilation gives a contribution
\be \Omega_\ax\sim \rfrac{t_\sub{ann}}{\tilde t} \threehalf
\ftw^{1.18} \gamma
\ee
where $t_\sub{ann}$ is the wall annihilation time. Thus,
axions emitted by the walls are probably
significant if the Harari-Sikivie hypothesis is correct, but
may not be if the Davis hypothesis is correct.

It seems difficult to go beyond these
rather crude considerations except through numerical simulations, which
after wall formation would become extremely difficult.

\subsection{The no-string scenario}

So far we have assumed that axionic strings are generated in the early
universe, either after inflation by the Kibble mechanism or during
inflation by the quantum fluctuation.
The criterion for the former possibility is
$T_\reh\gsim f_\ax$, and from Figure 3 one sees that this condition
is rather similar to the condition $T_\reh\gsim T_\decoup$
which corresponds to the presence of thermal saxinos in the universe.
The latter condition can hold without the former only if $T_\reh$ is
rather low, which means that the Kibble mechanism will certainly
generate strings if saxinos generate a significant amount of entropy.

The criterion for the generation of strings by the quantum fluctuation
is less certain. Considering only the usual non-supersymmetric
Peccei-Quinn field and ignoring any interaction with the inflaton field
or the spacetime curvature, the criterion was estimated in
\cite{lyst92} to be $H_1\gsim 1\times\fa$, where $H_1$ is the Hubble
parameter when the observable universe leaves the horizon.
This estimate should be fairly robust as a rough
criterion, and may perhaps be thought of as a quasi-thermal effect, due
to the Hawking temperature $H_1/2\pi$. It
implies that string production does not occur if $H_1<10^{10}\GeV$
(because $f_\ax>10^{10}\GeV$). While such a low value of $H_1$ is
possible \cite{lind91,lily93a}, most models of inflation require
$H_1\sim 10^{13}$ to $10^{14}\GeV$, in which case the no-string scenario
requires $f_a\gsim 10^{13}\GeV$ (Figure 3).

A rather complete treatment of the no-string scenario
has been given in \cite{lyth92b}, and we recall briefly the main
results, again ignoring any coupling of the Peccei-Quinn field to other
fields. The axion field is homogeneous
before the epoch $\tilde T$, except for an inhomogeneity which originates
as a vacuum fluctuation during inflation. Writing
$\tilde \theta
=\bar\theta+\delta\theta$, where $\bar\theta$ is the average of
$\tilde \theta$ in the observable universe,
\eq{27} for $\Omega_\ax$ becomes
\be
\Omega_\ax=0.9\times 10^{\pm.5} \ftw^{1.18} \Delta\mone (\bar\theta^2
+\sigma_\theta^2)
\label{55}
\ee
where $\sigma_\theta^2$, the average of
$(\delta\tilde\theta)^2$, is given by
\be
\sigma_\theta\simeq \frac{4}{2\pi} \frac{H_1}{\fa}
\label{56}
\ee
The inhomogeneity of the axion field causes a
primeval isocurvature density perturbation, with a flat spectrum given by
\cite{lyth92b}
\be
\calp_\sub{iso}\simeq \frac{\Omega_\ax^2}{16}
\frac{
4\bar\theta^2 \sigma_\theta^2+\sigma_\theta^4
}
{(\bar\theta^2+\sigma_\theta^2)^2}
\label{57}
\ee
(It might cause bags of domain wall as well
\cite{lily90}, but this possibility has yet to be fully explored.)
If present, a primeval isocurvature
density perturbation contributes to the large scale microwave background
anisotropy. Its effect on the rms anisotropy is the same
as that of an adiabatic density perturbation whose spectrum $\delta_H^2$
at horizon entry is given by
\be
\delta_H=6\times(2/15) \calp\half_\sub{iso}
\label{58}
\ee
The COBE measurement implies \cite{lily93a}
that $\delta_H=(1.7\pm0.3)\times 10^{-5}$ which corresponds to\footnote
{A weaker constraint $\calp_\sub{iso}\half<10^{-4}$ was used, corresponding
to the use of pre-COBE quadrupole data, but the difference is not very
significant for the present purpose.}
\be \calp\half_\sub{iso}<1.7\times 10^{-5}
\label{59}
\ee

At fixed $\bar\theta$, \eqss{55}{57}{59} forbid a
substantial region of the $m_\ax$--$H_1$ plane \cite{lyst92}.
 The forbidden region is
minimised for $\bar\theta=0$, but there is no reason to expect
$\bar\theta$ to be very small, unless a very small value
is demanded by the requirement $\Omega_\ax<1$. Suppose that
one sets
$\bar\theta=\min(.1\pi,\bar\theta_\Omega)$ where $\bar\theta_\Omega$ is
the value of $\bar\theta$ which makes $\Omega_\ax=1$,
and fixes $H_1$. Then an interval of $m_\ax$ is forbidden, which
covers the entire regime $H_1<f_\ax<\mpl$ if $H_1>10^{12}\GeV$
but which shrinks to nothing for $H_1<10^{10}\GeV$.

The right hand column of Figure 2 illustrates the case $H_1<10^{10}\GeV$;
to the left of the diagonal line
there is no constraint on $m_\ax$ except the astrophysical one.
As $H_1$ is increased from $10^{10}\GeV$ to $10^{12}
\GeV$ a forbidden region
rapidly develops and the situation  becomes practically the same as
for the string scenario.

\section{Conclusions}

The calculation presented here
is valid under the assumption that the correct extension of the Standard
Model respects Peccei-Quinn symmetry and supersymmetry,
the latter being implemented without introducing dramatic new features
such as a very light gravitino \cite{chki93}. Subject to this
requirement, the calculation shows that saxinos of thermal origin
can dilute estimates of cosmological mass densities
by a factor of up to $10^4$, and non-thermal saxinos might cause even
more dilution. Let us close by briefly
considering the implication of a dilution factor for some commonly
discussed cases \cite{kotu90},
leaving aside the axion which has already been treated.

The baryon density is usually considered these days to be generated at
the electroweak transition, and at least the more conservative proposals
do not generate significantly more than the observed density
(eg. \cite{fash93}). Accordingly these models could not tolerate a
large dilution factor, but as the theoretical situation is still very
fluid it is probably too early to say anything very definite.

Apart from the axion a favourite dark matter candidate is a
massive neutrino species. Species with mass $m_\nu\lsim 10\MeV$ decouple
while still relativistic, and without any dilution factor
$\Omega<1$ requires $m_\nu<100\eV$. Decoupling occurs
just before nucleosynthesis,
so this bound is not affected by entropy production
(which must finish by that epoch).
Species with mass $m_\nu\gsim 10\MeV$ on the other hand decouple
while non-relativistic, and without dilution this opens up an
additional allowed interval $m_\nu\gsim 1\GeV$. In that case
decoupling occurs at $70\MeV(m_\nu/1\GeV)$ \cite{kotu90}
so dilution can occur and it it does
the allowed interval extends to lower masses.
 From Fig.~5.2 of \cite{kotu90} one learns
that with a dilution factor $\Delta\sim 10^4$ the interval extends right
down to $m_\nu\sim10\MeV$.

Finally, consider a particle species with much weaker
interactions than the neutrino, so that decoupling occurs
at $T\gg 1\GeV$ when $g_*\sim100$. Without entropy dilution
the mass needed to make $\Omega=1$ is $m\sim 1\keV$,
 but with dilution the mass is
increased by a factor $\Delta$ and so could be of order
$10\MeV$. Such a particle species would constitute cold dark matter,
instead of `warm' dark matter as in the usual case.

{}From these examples one sees that a significant entropy dilution factor could
have profound implications.
Whether or not such a factor is generated by thermal saxinos
depends mainly on whether the reheat temperature is
high or low,  and we are reminded that an understanding of
this quantity is one of the outstanding problems of theoretical
cosmology.

\section{Acknowledgments}

I thank Ewan Stewart for collaboration in the early stages of this work,
and Rick Davis, Warren Perkins, Paul Shellard, Pierre Sikivie
and David Wands for useful conversations.

\subsection*{Figure Captions}

FIG.~1.---Plots against axion mass.
All scales are logarithims to base 10, and in
each plot the horizontal axis is $m_\ax/1\eV$. The rows
correspond to different saxino masses as indicated. The left hand
column shows various temperatures in units of $1\MeV$, the full line
indicating the temperature $\tilde T$ after which axion number is
conserved, and the dotted lines indicating
the three temperatures $T_\decay<T_=<T\sub{saxeq}$ defined in the text
(entropy generation occurs only in the narrow interval $T_\decay
<T<T_=$). The second column gives the entropy production factor $\Delta$,
given by \eq{18a} or (dotted line) \eq{17a}. The two right hand columns
give the axion density $\Omega_\ax$, for $T_\reh=10^{16}\GeV$ (full
lines) and $T_\reh<10^{8}\GeV$ (dotted lines and their continuation).
In each case a band of
values is shown to take into account the uncertainties discussed in the
text, and the second case is identical with the no-supersymmetry result.
The hatched regions correspond to the nucleosynthesis constraint and
(for the two right hand columns) the constraint $\Omega_\ax<1$.

FIG.~2.---The axion window. In each plot the vertical axis is
$\log_{10}(m_\ax/1\eV)$ and the horizontal axis is $\log_{10}
(T_\reh/1\GeV)$. The first row corresponds to the non-existence of the saxino
and the others to different saxino masses.
The three columns correspond to different possibilities
described in the text. The upper cross hatched region is forbidden
by accelerator physics and astrophysics. The lower cross hatched region
is forbidden either by $\Omega_\ax<1$ or by the requirement that
saxino decay must not interfere with nucleosynthesis.
The latter requirement is relevant only for the second and third rows,
giving the right hand horizontal part of the
boundary in each plot. In each plot (except on the first row)
the diagonal full line and its straight continuation
is the line below which thermal
saxinos are absent, so that there is no dilution factor
(non-thermal saxinos are not considered in the present paper).
The diagonal dotted line in the right hand column is the line below
which the Kibble mechanism does not produce axionic strings.

FIG.~3.---Various regimes of the $m_\ax$--$T_\reh$ plane, as
discussed in the text. The horizontal line marks the regime
$H_1>f_\ax$ for the case $H_1=10^{13}\GeV$, and it moves up one decade
for every decade that $H_1$ moves down.


\begin{thebibliography}{999}
\bibitem{kotu90} E W Kolb and M S Turner,
{\sl The Early Universe} (Addison-Wesley 1990).

\bibitem{hoet92}
R Holman {\em et al}, Phys Lett B 282, 132 (1992).
L Randall, Phys Lett B 284, 77 (1992).
K S Babu and S M Barr, Phys Lett b 300, 367 (1993).
R Holman, T W Kephart and Soo-Jong Rey, preprint (1992).

\bibitem{kama92}
M Kamionkowski and J March-Russel, Phys Lett B 282, 137 (1992).
S M Barr and D Seckel, Phys Rev D 46, 539 (1992).
S Ghigna, M Lusignoli and M Roncadelli, Phys Lett B 283, 278 (1992).
J Garcia-Bellido, preprint (1992).

\bibitem{pequ77} R D Peccei and H Quinn, Phys Rev Lett 38, 1440 (1977).

\bibitem{kim87} J E Kim, Phys Rep 150 1 (1987).
H Y Cheng, Phys Rep 158 1 (1988).

\bibitem{wein78} S Weinberg, Phys Rev Lett 40, 223 (1978).
F Wilczek, Phys Rev Lett 40, 279 (1978).

\bibitem{lind90} A D Linde, {\sl Particle Physics and Cosmology}
(Gordon and Breach, 1990).

\bibitem{wilc93} See, for instance, the rapporteur talks of
F Wilcjek and R Peccei at the 1992 TEXAS/ESO conference
(Berkeley), to be published
in the Conference Proceedings.

\bibitem{nill84} H P Nilles, Phys Rep 110, 1 (1984).

\bibitem{kim84} J E Kim, Phys Lett 136B, 378 (1984).

\bibitem{first} S A Bonometto, F Gabbiani and A Masiero, Phys Lett B222,
433 (1989).
K Rajagopal, M S Turner and F Wilczek, Nuc Phys B358,
447 (1991).
J E Kim and H P Nilles, Phys Lett B 263, 79 (1991).
E J Chun, J E Kim and H P Nilles, Nuc Phys B 370, 105 (1992).
E J Chun, J E Kim and H P Nilles, Phys Lett B287, 123 (1992).
S Mollerach and E Roulet, Phys Lett B281, 303 (1992).
S A Bonometto, F Gabbiani and A Masiero, Milan preprint (1993).

\bibitem{kim91} J E Kim, Phys Rev Lett 67, 3465 (1991).

\bibitem{chki93} E J Chun, H B Kim and J E Kim, preprint (1993).

\bibitem{amet91} U Amaldi, W de Boer and H Fursten,
Phys Lett B260, 477 (1991).

\bibitem{lyth84} L F Abbot and M B Wise, Nuc Phys B 244, 541 (1984).
D H Lyth, Phys Lett   147B  403 (1984); erratum
  150B  465(E) (1985).

\bibitem{lily93a} A R Liddle and D H Lyth, Phys Rep, to appear (1993).

\bibitem{kasa87} M Kawashi and K Sato, Phys Lett B189, 23 (1987).

\bibitem{lapa86}
G Lazarides, C Panagiotakopoulos and Q Shafi,
Phys Rev Lett 56, 432 (1986) and Phys Rev Lett 56, 557 (1986).
K Yamamoto, Phys Lett B168, 341 (1986).

\bibitem{lasc90} G Lazarides, C Panagiotakopoulos and Q Shafi,
Phys Lett B 192, 323 (1987).
G Lazarides, R Schaefer,
D Seckel and Q Shafi, Nuc Phys B346, 193 (1990).

\bibitem{sctu85} R J Scherrer and M S Turner, Phys Rev D31, 681 (1985).
\bibitem{sctu88} R J Scherrer and M S Turner, Astroph Journ
331, 19 (1988).

\bibitem{lyth90} D H Lyth, Phys Lett B 236, 408 (1990).

\bibitem{lily90} A D Linde and D H Lyth, Phys Lett  B246,  353 (1990).

\bibitem{tuwi91} M S Turner and F Wilcjek, Phys Rev Lett 66 5 (1991).

\bibitem{lind91} A Linde, Phys Lett B 259, 38 (1991).

\bibitem{lyth92a} D H Lyth, Phys Lett B275, 279 (1992).

\bibitem{lyth92b} D H Lyth, Phys Rev D45, 3394 (1992).

\bibitem{lyst92} D H Lyth and E D Stewart, Phys Lett B283, 189 (1992);
Phys Rev D46, 532 (1992).

\bibitem{davi86} R L Davis Phys Lett B  180  225 (1986).

\bibitem{hasi87} D Harari and P Sikivie Phys Lett B195 361 (1987).
P Sikivie, in
\sl Proceedings of Nobel Symposium No 79:
The Birth and Evolution of Our Universe,
 \rm eds B Guftafsson, Y Nilsson and B Skagerstam
(WSPC, Singapore, 1991).

\bibitem{rafe90} G Raffelt, Phys Rep 198 1 (1990).
G Raffelt and D Seckel, Phys Rev Lett 67, 2605 (1991).

\bibitem{turn86} M S Turner, Phys Rev D33, 889 (1986).

\bibitem{absi83} L F Abbot and P Sikivie, Phys Lett 120B, 1983.

\bibitem{dash89} R L Davis and E P S Shellard, Nuc Phys  B324,  167 (1989).

\bibitem{hasi90}
C  Hagmann and P Sikivie, Nuc Phys 363, 247 (1991).

\bibitem{kila82} T W B Kibble, G Lazarides and Q Shafi, Phys Rev D
26, 435 (1982). A Vilenkin and A E Everett, Phys Rev Lett 48, 1867
(1982). T Vachaspati, A E Everett and A Vilenkin (1984).

\bibitem{fash93} G R Farrar and M E Shaposhnikov, Phys Rev Lett 70,
2833 (1993).

\end{thebibliography}
\end{document}